\documentclass[11pt]{article}

\usepackage{amsmath}
\usepackage{graphicx}
\usepackage{amsfonts}
\usepackage{rotating}
\usepackage{color}
\newtheorem{theo}{Theorem}

\newtheorem{lemma}{Lemma}
\newenvironment{proof}{\trivlist\item[]\textbf{Proof.\ }}
                      {\endtrivlist}

\begin{document}
\title{\bf\Large Inference of $R=P(Y<X)$ for two-parameter Rayleigh distribution based on progressively censored samples}
\vspace{-1cm}
\author
{Akram Kohansal\footnote{{Department of Statistics, Imam Khomeini International University, Qazvin, Iran. Email: \texttt{kohansal@sci.ikiu.ac.ir}}} , Saeid Rezakhah\footnote{Corresponding author: Faculty of Mathematics and Computer Science, Amirkabir University of Technology,
Tehran, Iran. Email: \texttt{rezakhah@aut.ac.ir}}}
\maketitle
\begin{abstract}
Based on  independent  progressively Type-II censored samples from two-parameter Rayleigh distributions with the same location parameter but different scale parameters,  the UMVUE and maximum likelihood estimator of $R=P(Y<X)$ are obtained.   Also the exact, asymptotic and bootstrap confidence intervals for $R$ are evaluated.  Using Gibbs {sampling,} the Bayes estimator  and  corresponding credible interval for $R$  are obtained too. Applying Monte Carlo {simulations,}  we compare the performances of the different estimation methods.  Finally 
we make use of  simulated data and two real data sets to   show  the competitive performance of our method. 
\end{abstract}

\section{Introduction}
\par
The two most popular censoring schemes which widely are used in practice are  Type-I and Type-II.  By Type-I  the test is terminated when a pre-determined time on test has been reached and by Type-II when a pre-chosen number of failures has been observed. None of these schemes allows the removal of active units during the experiment. Progressive censoring schemes are based on removing some units after each failure.   Progressive Type-II censoring scheme is a combination of the  Progressive  censoring schemes and the Type-II one, which  has been very popular in the last decade.  Suppose $N$ units are placed on a life test and the experimenter decides to follow the test up to the time of {$n$-th}  failure and to remove  $R_i$ units randomly from the surviving ones at {the}  time of $i$-th failure, $i=1,2,\cdots n$, where $n+R_1+\ldots+R_n=N$.
  Therefore, {the} progressive Type-II censoring scheme consists of $n$ failure, and $n$ successive censoring of sizes $R_1,\ldots,R_n$.   This scheme includes the conventional Type-II right censoring scheme when
$R_1=\ldots=R_{n-1}=0$ and $R_n=N-n$ and complete sampling scheme when $N=n$ and $R_1=\ldots=R_n=0$. For further details on progressive censoring schemes and relevant references, the reader is referred to the book by Balakrishnan and Aggarwala \cite{2fi}.
\par
In reliability analysis, a general problem of interest  is inference of the stress-strength parameter $R = P(Y<X)$. The stress $Y$ and the
 strength $X$ are treated as independent random variables. The system fails when the applied stress is greater than the strength.
 Estimation of the stress-strength parameter has received considerable attention in the statistical literature. These studies started with the pioneering work
 of Birnbaum \cite{4fi}.
   Since then many studies have been accomplished on the estimation and inference of the stress-strength parameter, from the frequentist  and Bayesian  { point of view,  by} imposing different classes of distributions. The monograph by Kotz et al. \cite{9fi} provided { a comprehensive  review} of the development
  of this model till 2003. { Further  recent work on the stress-strength model can be fined} in Kundu and Gupta \cite{11fi,19sh},
   Raqab and Kundu \cite{26sh}, Krishnamoorthy et al. \cite{10fi}, Raqab et al. \cite{15fi}, Kundu and Raqab \cite{20sh}, 
 Panahi and Asadi \cite{25sh}, Lio and Tsai \cite{applied} {and Babayi et al.} \cite{Babayi}.
\par
Based on progressively Type-II censored samples, this paper deals with inference for the stress-strength reliability $R=P(Y<X)$ when $X$ and $Y$ are two independent two-parameter Rayleigh distributions with different scale parameters but having the same location parameter. This distribution was originally proposed by Khan et al. \cite{21pro}. 
 Statistical inference about this distribution was studied by Dey et al. \cite{American}.   {In the rest of the paper,  a two-parameter Rayleigh distribution with the pdf (\ref{we})   is denoted by $tR(\mu,\lambda)$. }
\par
{In this paper, we  study the estimation }  of the stress-strength parameter
$ R=P(Y<X)$ when $X$ and $Y$ are independent two-parameter Rayleigh random variables,    with common location parameter $\mu$ and scale parameters $\lambda>0$ and $\alpha>0$, respectively. The probability density functions (pdfs) of $X$ and $Y$ for $x>\mu$ and $y>\mu$ are;
\begin{equation}\label{we}f(x)=2\lambda(x-\mu)e^{-\lambda(x-\mu)^2} \hspace{.75cm}\mbox{and} \hspace{.75cm}
f(y)=2\alpha(y-\mu)e^{-\alpha(y-\mu)^2}.\end{equation}
 {This study follows under progressive Type-II censoring scheme for samples of  both random variables.
The progressive censoring schemes for $X$ and $Y$ are denoted by  $\{N, n, R_1,  R_2,  \ldots, R_{n}\}$ and $\{M,m,S_1,S_2,\ldots, S_{m}\}$  respectively.   Thus the observed progressively censored samples  of $X$ and $Y$ are denoted by  $\{X_{1:n:N},\ldots,X_{n:n:N}\}$ and $\{Y_{1:m:M},\ldots,Y_{m:m:M}\}$ and we are 
 to estimate $R=P(Y<X)=\frac{\alpha}{\alpha+\lambda}$. For computing the maximum likelihood estimator (MLE) of $R$, one need to  solve a non-linear equation.} So we propose to use a simple iterative method to find the MLE of $R$. { Also the} uniformly minimum variance unbiased estimator (UMVUE) of $R$ is provided. Furthermore, { the exact,  asymptotic} and two bootstrap confidence intervals of $R$ are { obtained}. { In addition we obtain mean, variance and density function  of the posterior estimates of $R$ under independent  gamma priors of $\lambda$ and $\alpha$  and improper uniform prior of $\mu$, and approximate Bayes estimate of $R$ under square error loss function.}

\par
{This  paper is organized }as follows: In Section 2, we discuss the maximum likelihood estimator of $R$ {and  show that the MLE can be obtained by an iterative method.} The UMVUE of $R$ is derived in Section 3. The exact, asymptotic and two bootstrap confidence intervals of $R$ are presented in Section 4. Bayes estimate  and the associated credible interval are discussed in Section 5. Simulation results and data analysis are presented in Sections 6. Finally, we conclude the paper in Section 7.

\section{Maximum Likelihood Estimation of $R$}
Let $X\sim tR(\mu,\lambda)$ and $Y\sim tR(\mu,\alpha)$ be independent random variables. Then
\begin{align}\label{R}
R=P(Y<X)&=\int_{\mu}^{\infty}P(Y<X|X=x)P(X=x)dx\nonumber\\
&=\int_{\mu}^{\infty}\{1-e^{-\alpha(x-\mu)^2}\}2\lambda (x-\mu) e^{-\lambda(x-\mu)^2}dx\nonumber\\
&=\int_{\mu}^{\infty}2\lambda (x-\mu) e^{-\lambda(x-\mu)^2}dx - \int_{\mu}^{\infty} 2\lambda (x-\mu) e^{-(\lambda+\alpha)(x-\mu)^2}dx\nonumber\\
&=1-\frac{\lambda}{\alpha+\lambda}=\frac{\alpha}{\alpha+\lambda}.
\end{align}
\par  Our aim is to obtain the MLE of $R$ based on progressive Type-II censored data for both variables.  
 So  {we need to evaluate  the MLE of $\mu$, $\lambda$ and $\alpha$.   Let $\{X_{1:n:N},\ldots,X_{n:n:N}\}$ and $\{Y_{1:m:M},\ldots,Y_{m:m:M}\}$ be two progressively censored samples from $tR(\mu,\lambda)$ and $tR(\mu,\alpha)$ under the progressive censoring schemes $\{N,n,R_1,R_2,\ldots,R_{n}\}$ and  $\{M,m,S_1,S_2,\ldots,S_{m}\}$, respectively.
So, the likelihood function of $\mu$, $\lambda$ and $\alpha$ is given by}
$$L(\mu,\lambda,\alpha)=\bigg[c_1\prod_{i=1}^{n}f(x_{i})[1-F(x_{i})]^{R_i}\bigg]\times \bigg[c_2\prod_{j=1}^{m}f(y_{j})[1-F(y_{j})]^{S_j}\bigg],$$
where
$$c_1=N(N-R_1-1)\cdots(N-R_1-\cdots-R_{n-1}-n+1),$$
$$c_2=M(M-S_1-1)\cdots(M-S_1-\cdots-S_{m-1}-m+1).$$
{Based on the observed data, the likelihood function is} \begin{align}\label{Ell}
L(data |\mu,\lambda,\alpha)&=c_1 c_2 (2\lambda)^n(2\alpha)^m\prod_{i=1}^{n} {(x_i-\mu)} \prod_{j=1}^{m} {(y_j-\mu)}\nonumber\\
&\times \exp\left[-\lambda\sum_{i=1}^{n}(R_i+1)(x_i-\mu)^2-\alpha\sum_{j=1}^{m}(S_j+1)(y_j-\mu)^2\right].\nonumber\\
\end{align}
So  the log-likelihood function  can be written as 
\begin{align*}
\ell(\mu,\lambda,\alpha)&=Constant+n\ln(\lambda)+m\ln(\alpha)+\sum_{i=1}^n\ln(x_i-\mu)+\sum_{j=1}^m\ln(y_j-\mu)\\
&-\left[\lambda\sum_{i=1}^{n}(R_i+1)(x_i-\mu)^2+\alpha\sum_{j=1}^{m}(S_j+1)(y_j-\mu)^2\right].
\end{align*}
Therefore,  the MLEs of $\lambda, \; \alpha$ and $\mu$  can be derived as the solution of 
\begin{align}\label{eqa}
\frac{\partial\ell}{\partial\lambda}&=\frac{n}{\lambda}-\sum_{i=1}^{n}(R_i+1)(x_i-\mu)^2=0,\\\
\label{eqb}
\frac{\partial\ell}{\partial\alpha}&=\frac{m}{\alpha}-\sum_{j=1}^{m}(S_j+1)(y_j-\mu)^2=0,\\\
\label{eqt}
\frac{\partial\ell}{\partial\mu}&=2\left[\lambda\sum_{i=1}^n(R_i+1)(x_i-\mu)+\alpha\sum_{j=1}^m(S_j+1)(y_j-\mu)\right]\nonumber\\
&\;\;\;\;-\sum_{i=1}^n\frac{1}{x_i-\mu}-\sum_{j=1}^m\frac{1}{y_j-\mu}=0.\nonumber
\end{align}
We denote these MLEs  by   $\hat\lambda$, $\hat\alpha$ and $\hat\mu$  respectively.  Thus 
\begin{align*}
\hat\lambda(\mu)=\frac{n}{\sum_{i=1}^{n}(R_i+1)(x_i-\mu)^2},\qquad
\hat\alpha(\mu)=\frac{m}{\sum_{j=1}^{m}(S_j+1)(y_j-\mu)^2},
\end{align*}
and $\hat\mu$ {is the solution of } $k(\mu)=\mu,$ where
\begin{align}
k(\mu)&=2\left[\frac{n\sum_{i=1}^n(R_i+1)(x_i-\mu)}{\sum_{i=1}^n(R_i+1)(x_i-\mu)^2}+\frac{m\sum_{j=1}^m(S_j+1)(y_j-\mu)}{\sum_{j=1}^m(S_i+1)(y_j-\mu)^2}\right]\nonumber\\
&\;\;\;\;-\sum_{i=1}^n\frac{1}{x_i-\mu}-\sum_{j=1}^m\frac{1}{y_j-\mu}+\mu.\nonumber
\end{align}
{By using an iterative scheme as
$k(\mu_{(j)}) = \mu_{(j+1)},$ where $\mu_{(j)}$ is the $j$-th iterate of $\hat\mu$, } which stops  when $|\mu_{(j)}-\mu_{(j+1)}|$ becomes sufficiently small.  Then   $\hat\mu$ is obtained  and  the values of  $\hat\lambda$ and $\hat\alpha$ are  resulted.  Therefore, the MLE of $R$ is evaluated as 
\begin{equation}\label{hat_R}
\hat R=\frac{\hat\alpha}{\hat\alpha+\hat\lambda}.
\end{equation}

\section{UMVUE of $R$}
{Let $\{X_{1:n:N},\ldots,X_{n:n:N}\}$ and $\{Y_{1:m:M},\ldots,Y_{m:m:M}\}$ be two progressively censored samples  from $tR(\mu,\lambda)$ and $tR(\mu,\alpha)$ and 
under the  schemes $\{N,n,R_1,R_2,\ldots,R_{n}\}$ and  $\{M,m,S_1,S_2,\ldots,S_{m}\}$,   respectively.} Thus the joint pdf of $X_{1:n:N},\ldots,X_{n:n:N}$ can be written as 
\begin{align}\label{ff}
f_{X_{1:n:N},\ldots,X_{n:n:N}}(x_1,\ldots,x_n)=c_1 (2\lambda)^n\prod_{i=1}^n{(x_i-\mu)}\exp\left[-\lambda\sum_{i=1}^n(R_i+1)(x_i-\mu)^2\right],
\end{align}
where $\mu<x_1<\ldots<x_n<\infty$. {The equation (\ref{ff}) indicates that} $U=\sum_{i=1}^n(R_i+1)(X_i-\mu)^2$ is the complete sufficient statistics for $\lambda$ when the location parameter $\mu$ is known.  One can easily verify that $X^*_{i:n:N}=(X_{i:n:N}-\mu)^2$ has an exponential distribution with mean $\lambda^{-1}$.   By applying  the transformations 
\begin{align*}
Z_1&=NX^*_{1:n:N},\\
Z_2&=(N-R_1-1)(X^*_{2:n:N}-X^*_{1:n:N}),\\
\vdots\\
Z_n&=(N-R_1-\ldots-R_{n-1}-n+1)(X^*_{n:n:N}-X^*_{n-1:n:N}), 
\end{align*}
{Cao and Cheng \cite{iei} showed that} $Z_1,\ldots,Z_n$ are independent and identically distributed exponential random variables with mean $\lambda^{-1}$.  Therefore,  $U=\sum_{i=1}^nZ_i=\sum_{i=1}^n(R_i+1)X^*_{i:n:N}$  { has gamma distribution with the following pdf:}
\begin{align}\label{fU}
f_U(u)=\frac{\lambda^n}{\Gamma(n)}u^{n-1}\exp(-\lambda u),~~u>0.
\end{align}

\begin{lemma}\label{lemnew}
The conditional pdf of $X^*_{1:n:N}$ given $U=\sum_{i=1}^n(R_i+1)X^*_{i:n:N}=u$ can be written as
$$f_{X^*_{1:n:N}|U=u}(x)=N(n-1)\frac{(u-Nx)^{n-2}}{u^{n-1}},~~0<x<u/n.$$
Also  by assuming    $Y^*_{j:m:M}=(Y_{j:m:M}-\mu)^2$ and $V=\sum_{j=1}^m(S_j+1)(Y_j-\mu)^2$, the conditional pdf of $Y^*_{1:m:M}$ given $V=v$ can be written as
$$f_{Y^*_{1:m:M}|V=v}(y)=M(m-1)\frac{(v-My)^{m-2}}{v^{m-1}},~~0<y<v/m.$$
\end{lemma}

 \proof { We have that $$f_{X^*_{1:n:N}|U=u}(x)=\frac{f_{X^*_{1:n:N},U}(x,u)}{f_U(u)},$$ where $f_{X^*_{1:n:N},U}(x,u)$ is the the joint pdf of $X^*_{1:n:N}$ and $U$ is the complete sufficient statistics for $\lambda$ when $\mu$ is known that  its  pdf is given by  (\ref{fU}).  Let  $W=\sum_{i=2}^n Z_i$.  So  $W$ and $Z_1$ are independent.   Using the transformation $Z_1=NX^*_{1:n:N}$,  $U=W+Z_1$, one can easily derive the joint pdf  of  $X^*_{1:n:N}$ and $U$ from  the joint pdf  of $W$ and $Z_1$.   Finally, using (\ref{fU}) the result is derived.} The second part of the theorem  follows by the same reasoning.  $\square$

\begin{theo}\label{theo3}
{The UMVUE of $R$,  $n,m \geq 2$,  is
\begin{equation}\label{UMVUE}
\tilde R=\left\{
\begin{array}{rl}
1-\sum\limits_{k=0}^{m-1}(-1)^k (\frac{u}{v})^k \frac{{{m-1}\choose k}}{{{n+k-1}\choose k}}& \text{if } u<v,\\
 \sum\limits_{k=0}^{n-1}(-1)^k (\frac{v}{u})^k \frac{{{n-1}\choose k}}{{{m+k-1}\choose k}}& \text{if } u>v,
\end{array} \right.
\end{equation}
where $U$ and $V$ are the  statistics  defined in Lemma 1.
}
\end{theo}

\proof $X^*_{1:n:N}\sim Exp(\frac{1}{N\lambda})$ and $Y^*_{1:m:M}\sim Exp(\frac{1}{M\alpha})$, so 
\begin{align*}
\phi(X^*_{1},Y^*_{1})=\left\{
\begin{array}{rl}
1& \text{if } MY^*_{1:m:M}<NX^*_{1:n:N},\\
 & \\
0& \text{if } MY^*_{1:m:M}>NX^*_{1:n:N},
\end{array} \right.
\end{align*}
is an unbiased estimator of $R$. Therefore,
\begin{align*}
\tilde R=E[\phi(X^*_{1},Y^*_{1})|U=u,V=v]=\iint_{\cal{A}}f_{X^*_{1}|U=u}(x)f_{Y^*_{1}|V=v}(y)dxdy,
\end{align*}
where ${\cal{A}}=\{(x,y):0<x<u/N,0<y<v/M,Nx>My\}.$   Also $f_{X^*_{1}|U=u}(x)$ and $f_{Y^*_{1}|V=v}(y)$ are
 defined in Lemma \ref{lemnew}.  Thus, for $u<v$
\begin{align*}
\tilde R&=\frac{N(n-1)}{u^{n-1}}\frac{M(m-1)}{v^{m-1}}\int_{0}^{u/N}\int_{0}^{Nx/M}(u-Nx)^{n-2}(v-My)^{m-2}dydx\\
&=1-\frac{N(n-1)}{u^{n-1}v^{m-1}}\int_{0}^{u/N}(u-Nx)^{n-2}(v-Nx)^{m-1}dx~~\{Put: \frac{Nx}{u}=t\}\\
&=1-(n-1)\int_0^1(1-t)^{n-2}(1-\frac{ut}{v})^{m-1}dt\\
&=1-(n-1)\int_0^1(1-t)^{n-2}\sum_{k=0}^{m-1}(-1)^k\binom{m-1}{k}(\frac{ut}{v})^kdt\\
&=1-\sum_{k=0}^{m-1}(-1)^{k}(\frac{u}{v})^k\frac{\binom{m-1}{k}}{\binom{n+k-1}{k}}.
\end{align*}
By the same method one can easily verify that for $u>v$, $\tilde R=\sum_{k=0}^{n-1}(-1)^{k}(\frac{v}{u})^k\frac{\binom{n-1}{k}}{\binom{m+k-1}{k}}$. $\square$

\section{Confidence Intervals }
\subsection{Exact confidence interval}
{In this subsection we obtain  an exact confidence interval for $R$. Let $\{X_{1:n:N},\ldots,X_{n:n:N}\}$ and $\{Y_{1:m:M},\ldots,Y_{m:m:M}\}$ be progressively censored samples from $tR(\mu,\lambda)$ and $tR(\mu,\alpha)$ under  censoring schemes $\{N,n,R_1,R_2,\ldots,R_{n}\}$ and  $\{M,m,S_1,S_2,\ldots,S_{m}\}$, respectively.} Let $X^{**}_{i:n:N}=\lambda(X_{i:n:N}-\mu)^2,~i=1,\cdots,n$ and  $Y^{**}_{j:m:M}=\alpha(Y_{j:m:M}-\mu)^2,~j=1,\cdots,m$.   So 
 $X^{**}_{i:n:N},~i=1,\cdots,n$  and  $Y^{**}_{j:m:M},~j=1,\cdots,m$   are  two progressively censored  samples from the  standard exponential distribution. Now consider the following transformations:
\begin{align*}
Z^*_1&=NX^{**}_{1:n:N},~Z^*_i=N-\sum_{k=1}^{i-1}(R_k+1)(X^{**}_{i:n:N}-X^{**}_{i-1:n:N}),~i=2,\cdots,n,\\
D^*_1&=MY^{**}_{1:m:M},~D^*_j=M-\sum_{k=1}^{j-1}(S_k+1)(Y^{**}_{j:m:M}-Y^{**}_{j-1:m:M}),~j=2,\cdots,m.
\end{align*}
By Cao and Cheng \cite{iei} we have  that $Z^*_1,\cdots,Z^*_n$  and  $D^*_1,\cdots,D^*_m$  are  two sets of  i.i.d.    standard exponential random variables.   
\par
Let $V_X=2Z^*_1$,   $U_X=2\sum_{i=2}^{n}Z^*_i$   and  also $V_Y=2D^*_1$ and $U_Y=2\sum_{j=2}^{m}D^*_j$. Then $V_X$ and $U_X$  are independent random variables  and so are $V_Y$ and $U_Y$.   Further 
$V_X\sim\chi^2_{(2)},~ ~U_X\sim\chi^2_{(2n-2)}$ { and also}  $~ V_Y\sim\chi^2_{(2)}~\mbox{and}~U_Y\sim\chi^2_{(2m-2)}).$

\begin{lemma}\label{lem1}
Let  $T_X(\mu)=\frac{U_X}{(n-1)V_X}$, $T_1=U_X+V_X$, $T_Y(\mu)=\frac{U_Y}{(m-1)V_Y}$ and $T_2=U_Y+V_Y$.   Then all these statistics are independent and  $$T_X(\mu)\sim F(2n-2,2),~T_1\sim\chi^2(2n), ~T_Y(\mu)\sim F(2m-2,2) ~\mbox{and}~T_2\sim\chi^2(2m).$$ 
\proof  
The distributions  of  all  statistics follows easily. By  the independence of two set of  samples, the independence of  
$T_X(\mu)$ and $T_1$ from  $T_Y(\mu)$ and $T_2$  follows.  By  Johnson et al. \cite{11first} 
  the independence  of $T_X(\mu)$  and  $T_1$ and of $T_Y(\mu)$  and   $T_2$ follows. $\square$
\end{lemma}

\begin{lemma}\label{lem2}
 $T_X(\mu)$  and $T_Y(\mu)$  are  strictly increasing in $\mu$.
\end{lemma}
\proof Let $\xi(\mu)=(\frac{a_i-\mu}{a_1-\mu})^2$ : $\mu<a_1<a_i~, i=2,\ldots,n$.
{  Then 
$\frac{d \xi(\mu)}{d\mu}=2\frac{(a_i-\mu)(a_i-a_1)}{(a_1-\mu)^3}>0$, so  $\xi(\mu)$ is strictly increasing in $\mu$.  
   Also, one can easily verify that  $T_X(\mu)=\frac{1}{N(n-1)}\sum_{i=1}^n(R_i+1)\left(\frac{X_i-\mu}{X_1-\mu}\right)^2-\frac{1}{n-1}.$}
Hence, $T_X(\mu)$ is  strictly increasing in $\mu$. $\square$

\begin{lemma}\label{lem3}
When $\mu$ is known, the MLE of $R$ {can be written as}
\begin{equation}\label{R_known}
\hat R_{ML}=\frac{1}{1+\frac{nT_2/\alpha}{mT_1/\lambda}},
\end{equation}
where   $T_1$ and $T_2$ are defined by Lemma \ref{lem1}.  Also $\hat R_{ML}=\frac{1}{1+\frac{\lambda}{\alpha}F}$ and $F=\frac{R}{1-R}.\frac{1-\hat R}{\hat R}$,
  where $F\sim F(2m,2n)$.
\proof
By \ref{hat_R} and  Lemma  \ref{lem1} we have \ref{R_known}.  Also by Lemma \ref{lem1}  $T_1$ and $T_2$ are independent with the presented  chi-square distributions. So    $\hat R_{ML}=\frac{1}{1+\frac{\lambda}{\alpha}F}$ 
 and $F=\frac{R}{1-R}.\frac{1-\hat R}{\hat R}$,
 where $F\sim F(2m,2n)$. $\square$
\end{lemma}

\begin{theo}
Suppose that {$\{X_{1:n:N},\ldots,X_{n:n:N}\}$ and $\{Y_{1:m:M},\ldots,Y_{m:m:M}\}$  are two progressively censored samples from $tR(\mu,\lambda)$ and $tR(\mu,\alpha)$, respectively.} Then
\begin{description}
\item[(i)] for any $0<\eta<1$,
\begin{align*}
\Big(\max\left\{T_X^{-1}\right.&\left.(F_{(1-\sqrt{1-\eta})/2}(2n-2,2)),T_Y^{-1}(F_{(1-\sqrt{1-\eta})/2}(2m-2,2))\right\},\\
&\min\left\{T_X^{-1}(F_{(1+\sqrt{1-\eta})/2}(2n-2,2)),T_Y^{-1}(F_{(1+\sqrt{1-\eta})/2}(2m-2,2))\right\}\Big)
\end{align*}
is a $100(1-\eta)\%$ confidence interval for $\mu$, where $F_{\eta}(p,q)$ is 100$\eta$-th percentile of $F(p,q)$.
\item[(ii)] for any $0<\eta<1$, the following inequalities determine a $100(1-\eta)\%$ joint confidence region for $(\mu,R)$,
\begin{equation*}
\hspace{-1cm}\left\{
\begin{array}{ll}
\max\left\{T_X^{-1}(F_{(1-\sqrt[4]{1-\eta})/2}(2n-2,2)),T_Y^{-1}(F_{(1-\sqrt[4]{1-\eta})/2}(2m-2,2))\right\}\leq\mu\\\
\hspace{.5in}\leq\min\left\{T_X^{-1}(F_{(1+\sqrt[4]{1-\eta})/2}(2n-2,2)),T_Y^{-1}(F_{(1+\sqrt[4]{1-\eta})/2}(2m-2,2))\right\},
\\ \\
\frac{1}{1+\frac{1-\hat R_{ML}}{\hat R_{ML}}F_{1-(1-\sqrt{1-\eta})/2}(2n,2m)}\leq R\leq\frac{1}{1+\frac{1-\hat R_{ML}}{\hat R_{ML}}F_{1-(1+\sqrt{1-\eta})/2}(2n,2m)}.
\end{array} \right.
\end{equation*}
\end{description}
\end{theo}
\proof \begin{description}
\item[(i)] By using Lemma \ref{lem1} and \ref{lem2}, we have
\begin{align*}
1-\eta&=\sqrt{1-\eta}.\sqrt{1-\eta}\\
&=P\Big[F_{(1-\sqrt{1-\eta})/2}(2n-2,2)\leq T_X(\mu)\leq F_{(1+\sqrt{1-\eta})/2}(2n-2,2)\Big]\\
&\hspace{0.5in}\times P\Big[F_{(1-\sqrt{1-\eta})/2}(2m-2,2)\leq T_Y(\mu)\leq F_{(1+\sqrt{1-\eta})/2}(2m-2,2)\Big]\\
&=P\Big[F_{(1-\sqrt{1-\eta})/2}(2n-2,2)\leq T_X(\mu)\leq F_{(1+\sqrt{1-\eta})/2}(2n-2,2),\\\
&\hspace{0.5in}F_{(1-\sqrt{1-\eta})/2}(2m-2,2)\leq T_Y(\mu)\leq F_{(1+\sqrt{1-\eta})/2}(2m-2,2)\Big]\\
&=P\Big[T_X^{-1}(F_{(1-\sqrt{1-\eta})/2}(2n-2,2))\leq \mu\leq T_X^{-1}(F_{(1+\sqrt{1-\eta})/2}(2n-2,2)),\\\
&\hspace{0.5in}T_Y^{-1}(F_{(1-\sqrt{1-\eta})/2}(2m-2,2))\leq \mu\leq T_Y^{-1}(F_{(1+\sqrt{1-\eta})/2}(2m-2,2))\Big]\\
&=P\Big[\max\left\{T_X^{-1}(F_{(1-\sqrt{1-\eta})/2}(2n-2,2)),T_Y^{-1}(F_{(1-\sqrt{1-\eta})/2}(2m-2,2))\right\}\leq \mu\\\
&\hspace{0.5in}\leq\min\left\{T_X^{-1}(F_{(1+\sqrt{1-\eta})/2}(2n-2,2)),T_Y^{-1}(F_{(1+\sqrt{1-\eta})/2}(2m-2,2))\right\}\Big].
\end{align*}
\item[(ii)] By using Lemma \ref{lem2} and Lemma \ref{lem3}, we have
\begin{align*}
&\hspace{-0.3in}P\Big[\max\left\{T_X^{-1}(F_{(1-\sqrt[4]{1-\eta})/2}(2n-2,2)),T_Y^{-1}(F_{(1-\sqrt[4]{1-\eta})/2}(2m-2,2))\right\}\leq \mu\\\
&\leq\min\left\{T_X^{-1}(F_{(1+\sqrt[4]{1-\eta})/2}(2n-2,2)),T_Y^{-1}(F_{(1+\sqrt[4]{1-\eta})/2}(2m-2,2))\right\},\\\
&\frac{1}{1+\frac{1-\hat R_{ML}}{\hat R_{ML}}F_{1-(1-\sqrt{1-\eta})/2}(2n,2m)}\leq R\leq\frac{1}{1+\frac{1-\hat R_{ML}}{\hat R_{ML}}F_{1-(1+\sqrt{1-\eta})/2}(2n,2m)}\Big]\\
&\hspace{-0.3in}=P\Big[T_X^{-1}(F_{(1-\sqrt[4]{1-\eta})/2}(2n-2,2))\leq \mu\leq T_X^{-1}(F_{(1+\sqrt[4]{1-\eta})/2}(2n-2,2)),\\\
&\hspace{0.1in}T_Y^{-1}(F_{(1-\sqrt[4]{1-\eta})/2}(2m-2,2))\leq \mu\leq T_Y^{-1}(F_{(1+\sqrt[4]{1-\eta})/2}(2m-2,2)),\\\
&\hspace{0.1in}F_{1-(1+\sqrt{1-\eta})/2}(2n,2m)\leq\frac{1-R}{R}.\frac{\hat R_{ML}}{1-\hat R_{ML}}\leq F_{1-(1-\sqrt{1-\eta})/2}(2n,2m)\Big]\\
&\hspace{-0.3in}=P\Big[F_{(1-\sqrt[4]{1-\eta})/2}(2n-2,2)\leq T_X(\mu)\leq F_{(1+\sqrt[4]{1-\eta})/2}(2n-2,2),\\\
&\hspace{0.1in}F_{(1-\sqrt[4]{1-\eta})/2}(2m-2,2)\leq T_Y(\mu)\leq F_{(1+\sqrt[4]{1-\eta})/2}(2m-2,2),\\\
&\hspace{0.1in}F_{(1-\sqrt{1-\eta})/2}(2m,2n)\leq\frac{R}{1-R}.\frac{1-\hat R_{ML}}{\hat R_{ML}}\leq F_{(1+\sqrt{1-\eta})/2}(2m,2n)\Big]\\
&\hspace{-0.3in}=P\Big[F_{(1-\sqrt[4]{1-\eta})/2}(2n-2,2)\leq T_X(\mu)\leq F_{(1+\sqrt[4]{1-\eta})/2}(2n-2,2)\Big]\\\
&\hspace{0.1in}\times P\Big[F_{(1-\sqrt[4]{1-\eta})/2}(2m-2,2)\leq T_Y(\mu)\leq F_{(1+\sqrt[4]{1-\eta})/2}(2m-2,2)\Big]\\\
&\hspace{0.1in}\times P\Big[F_{(1-\sqrt{1-\eta})/2}(2m,2n)\leq\frac{R}{1-R}.\frac{1-\hat R_{ML}}{\hat R_{ML}}\leq F_{(1+\sqrt{1-\eta})/2}(2m,2n)\Big]\\
&\hspace{-0.3in}=\sqrt[4]{1-\eta}.\sqrt[4]{1-\eta}.\sqrt{1-\eta}=1-\eta. ~\square
\end{align*}
\end{description}

\subsection{Asymptotic confidence interval}
The asymptotic distribution of the MLEs  $\hat\lambda$, $\hat\alpha$ and $\hat\mu$  are given by the entries of the inverse of the Fisher information matrix $J_{ij}=E\{-\partial^2\ell/\partial\theta_i\partial\theta_j\}$, where $i, j=1,2,3$ and $\Theta=(\theta_1,\theta_2,\theta_3)=(\lambda,\alpha,\mu)$. If the random variable $X$ follows two-parameter Rayleigh distribution as in (\ref{we}), then all the elements of the expected Fisher information matrix are not finite {(see Dey et al. \cite{American})}. Therefore, the ij-th element of  observed Fisher information matrix  is considerred as  $I_{ij}=\{-\partial^2\ell/\partial\theta_i\partial\theta_j\}_{\Theta=\hat\Theta}$,
 which is obtained by dropping the expectation operator $E$. The elements of the  observed Fisher information matrix has second partial derivatives of log-likelihood function as the entries, which can be obtained as follows:
\begin{align*}
I_{11}&=-\frac{\partial^2\ell}{\partial\lambda^2}=\frac{n}{\lambda^2},\qquad
I_{22}=-\frac{\partial^2\ell}{\partial\alpha^2}=\frac{m}{\alpha^2},\qquad\;\;\;
I_{12}=\frac{\partial^2\ell}{\partial\lambda\partial\alpha}=0=I_{21},\\
I_{13}&=-\frac{\partial^2\ell}{\partial\lambda\partial\mu}=-2\sum_{i=1}^{n}(R_i+1)(x_{i}-\mu)=I_{31},\\
I_{23}&=-\frac{\partial^2\ell}{\partial\alpha\partial\mu}=-2\sum_{j=1}^{n}(S_j+1)(y_{j}-\mu)=I_{32},\\
I_{33}&=-\frac{\partial^2\ell}{\partial\mu^2}=2\left[\lambda\sum_{i=1}^{n}(R_i+1)+\alpha\sum_{j=1}^{n}(S_j+1)\right]+\sum_{i=1}^{n}\frac{1}{(x_{i}-\mu)^2}+\sum_{j=1}^{m}\frac{1}{(y_{j}-\mu)^2}.\nonumber
\end{align*}

\begin{theo}\label{theon1}
As $n\rightarrow\infty,$ $m\rightarrow\infty$, and $n/m\rightarrow p$ then\\
$$[\sqrt n(\hat\lambda-\lambda)~~\sqrt n(\hat\alpha-\alpha)~~\sqrt m(\hat\mu-\mu)]^T\stackrel{\;\;\;D}{\longrightarrow} N_3(0,\mathbf{A^{-1}}(\lambda,\alpha,\mu)),$$
where $\mathbf{A(\lambda,\alpha,\mu)}$ and $\mathbf{A^{-1}(\lambda,\alpha,\mu)}$ are symmetric matrices as
\begin{equation*}
\mathbf{A(\lambda,\alpha,\mu)} = \left(
\begin{array}{ccc}
\frac{I_{11}}{n} & 0                       & \frac{I_{13}}{\sqrt{nm}} \\
                         & \frac{I_{22}}{n} & \frac{I_{23}}{\sqrt{nm}} \\
                         &                           & \frac{I_{33}}{m}
\end{array} \right),~~
\mathbf{A^{-1}(\lambda,\alpha,\mu)} = \frac{1}{|\mathbf{A(\lambda,\alpha,\mu)}|}\left(
\begin{array}{ccc}
\frac{b_{11}}{nm} & \frac{b_{12}}{nm} & \frac{b_{13}}{n\sqrt{nm}} \\
                             & \frac{b_{22}}{nm} & \frac{b_{23}}{n\sqrt{nm}} \\
                             &                              & \frac{b_{33}}{n^2}
\end{array} \right),
\end{equation*}
in which   $|\mathbf{A(\lambda,\alpha,\mu)}|=\frac{1}{n^2m}\left(I_{11}I_{22}I_{33}-I_{11}I_{23}^2-I_{13}^2I_{22}\right),$
\begin{align}
b_{11}&=I_{22}I_{33}-I^{2}_{23},\hspace{.5cm}
b_{12}=I_{13}I_{23},\hspace{.5cm}
b_{13}=-I_{13}I_{22},\nonumber\\
b_{22}&=I_{11}I_{33}-I^{2}_{13},\hspace{.5cm}
b_{23}=-I_{11}I_{23},\hspace{.5cm}
b_{33}=I_{11}I_{22}.\nonumber
\end{align}
\end{theo}

\proof Using the asymptotic properties of MLEs and the multivariate central limit theorem, we have that 
$$[(\hat\lambda-\lambda)~~(\hat\alpha-\alpha)~~(\hat\mu-\mu)]^T\stackrel{\;\;\;D}{\longrightarrow}  N_3(0,\mathbf{I^{-1}(\lambda,\alpha,\mu)}),$$
where $\mathbf{I(\lambda,\alpha,\mu)}$ and $\mathbf{I^{-1}(\lambda,\alpha,\mu)}$ are symmetric matrices as 
\begin{equation*}
\mathbf{I(\lambda,\alpha,\mu)} = \left(
\begin{array}{ccc}
I_{11} & 0         & I_{13} \\
           & I_{22} & I_{23} \\
           &             & I_{33}
\end{array} \right),~~
\mathbf{I^{-1}(\lambda,\alpha,\mu)} = \frac{1}{|\mathbf{I(\lambda,\alpha,\mu)}|}\left(
\begin{array}{ccc}
b_{11} & b_{12} & b_{13} \\
            & b_{22} & b_{23} \\
            &              & b_{33}
\end{array} \right),
\end{equation*}
and $|\mathbf{I(\lambda,\alpha,\mu)}|=n^2m|\mathbf{A(\lambda,\alpha,\mu)}|.$ Let
$
\mathbf{C} = \left(
\begin{array}{ccc}
\sqrt{n} & 0           & 0 \\
   0        & \sqrt{n} & 0 \\
   0        &  0          & \sqrt{m}
\end{array} \right)
$
then $$\mathbf{C}[(\hat\lambda-\lambda)~~(\hat\alpha-\alpha)~~(\hat\mu-\mu)]^T\stackrel{\;\;\;D}{\longrightarrow}  N_3(0,\mathbf{CI^{-1}(\lambda,\alpha,\mu)C^T}),$$
where $\mathbf{C}[(\hat\lambda-\lambda)~~(\hat\alpha-\alpha)~~(\hat\mu-\mu)]^T=[\sqrt n(\hat\lambda-\lambda)~~\sqrt n(\hat\alpha-\alpha)~~\sqrt m(\hat\mu-\mu)]^T$ and
$$\mathbf{CI^{-1}(\lambda,\alpha,\mu)C^T} = \frac{1}{|\mathbf{A(\lambda,\alpha,\mu)}|}\left(
\begin{array}{ccc}
\frac{b_{11}}{nm} & \frac{b_{12}}{nm} & \frac{b_{13}}{n\sqrt{nm}} \\
                             & \frac{b_{22}}{nm} & \frac{b_{23}}{n\sqrt{nm}} \\
                             &                              & \frac{b_{33}}{n^2}
\end{array} \right).$$
Therefore, the result follows. $\square$

\begin{theo}\label{theon2}
Let  $n,m \rightarrow\infty$  and $n/m\rightarrow p$.  Then\\
\begin{align}
\sqrt n(\hat R-R)\stackrel{\;\;\;D}{\longrightarrow} N(0,B),\nonumber
\end{align}
where
$$B=\frac{1}{(nm)|\mathbf{A(\lambda,\alpha,\mu)}|(\lambda+\alpha)^4}\bigg[\lambda^2 {b_{22}}+\alpha^2{b_{11}}-2\lambda\alpha {b_{12}}\bigg].
$$
\end{theo}
\proof Using Theorem \ref{theon1} and applying delta method,  we derive   the asymptotic distribution of $\hat R= g(\hat\lambda, \hat\alpha, \hat\mu)$ as
\begin{align}
\sqrt n(\hat R-R)\rightarrow N(0,B),\nonumber
\end{align}
where $B=\mathbf{b^TA^{-1}(\lambda,\alpha,\mu)b}$ in which  $\mathbf{b} = [\frac{\partial g}{\partial\lambda}~~\frac{\partial g}{\partial\alpha} ~~ \frac{\partial g}{\partial\mu}]^T=\frac{1}{(\lambda+\alpha)^2}[-\alpha ~~\lambda~~ 0]^T$ and $\mathbf{A^{-1}(\lambda,\alpha,\mu)}$ is defined in Theorem \ref{theon1}. Therefore, 
$$B=\mathbf{b^TA^{-1}(\lambda,\alpha,\mu)b}=\frac{1}{(nm)|\mathbf{A(\lambda,\alpha,\mu)}|(\lambda+\alpha)^4}\bigg[\lambda^2 {b_{22}}+\alpha^2{b_{11}}-2\lambda\alpha {b_{12}}\bigg].$$
Thus the Theorem is proved. $\square$\\ 

By Theorem \ref{theon2}, {  we have the    $100(1-\eta)\%$ asymptotic confidence interval for $R$  as }
$$(\hat R-z_{1-\frac{\eta}{2}}\frac{\sqrt{\hat B}}{\sqrt{n}},\hat R+z_{1-\frac{\eta}{2}}\frac{\sqrt{\hat B}}{\sqrt{n}}),$$
where $z_{\eta}$ is 100$\eta$-th percentile of $N(0,1)$.

\subsection{Confidence interval based on bootstrap procedures}
{If the parameter $\mu$ is unknown, the  distribution of $\hat R$ is not available and the asymptotic confidence interval can not be evaluated.   Also for small samples, the  asymptotic confidence intervals do not perform well.  Bootstrap method is another way to provide an approximate confidence interval for  $R$  in such a situations.   
 We present  two confidence intervals based on the non-parametric bootstrap methods as}  bootstrap-p ( Boot-p) method, based on the idea of Efron \cite{10sh}, and  bootstrap-t(Boot-t) method, based on the idea of Hall \cite{15sh}.
\\\\
\textbf{(i) Boot-p Method: }
This method is based on the following three steps.
\begin{itemize}
\item 1. {Generate two bootstrap samples $\{x_1^*,\ldots,x_{n}^*\}$ and $\{y_1^*,\ldots,y_{m}^*\}$ from $\{x_1,\ldots,x_{n}\}$ and $\{y_1,\ldots,y_{m}\}$, respectively. Compute $\hat R^*$, the bootstrap estimate of $R$  by  (\ref{hat_R}) and based on bootstrap samples  $\{x_1^*,\ldots,x_{n}^*\}$ and $\{y_1^*,\dots,y_{m}^*\}$.}
\item 2. Repeat step 1  NBOOT times.
\item 3. {Define $\hat R_{Bp}(x) = {G^{*}}^{-1}(x)$, where $G^*(x) = P(\hat R^*\leq x)$ is the empirical   cumulative distribution function of $\hat R^*$. The approximate $100(1-\eta)\%$ Boot-p confidence interval of $R$ can be written as}
\begin{equation}
(\hat R_{Bp}(\frac{\eta}{2}),\hat R_{Bp}(1-\frac{\eta}{2})).\nonumber
\end{equation}
\end{itemize}
\textbf{(ii) Boot-t Method: }
This  method is implemented by the following steps. 
\begin{itemize}
\item 1. {Compute $\hat R$ from the samples $\{x_1,\ldots,x_{n}\}$ and  $\{y_1,\ldots,y_{m}\}$.
\item 2. Generate two bootstrap samples $\{x_1^*,\ldots,x_{n}^*\}$ and $\{y_1^*,\ldots,y_{m}^*\}$ from $\{x_1,\ldots,x_{n}\}$ and $\{y_1,\ldots,y_{m}\}$, respectively. Compute $\hat R^*$, the bootstrap estimate of $R$, and the statistics
$$T^*=\frac{\sqrt n(\hat R^*-\hat R)}{\sqrt{V(\hat R^*)}},$$
where $V(\hat R^*)$, the asymptotic  variance  of $\hat R^*$, is obtained by Theorem \ref{theon2}.}
\item 3. Repeat steps 1 and 2 NBOOT times.
\item 4. {Define $\hat R_{Bt}(x)= \hat R + n^{-\frac{1}{2}}H^{-1}(x)\sqrt {V(\hat R )}$, where $H(x)=P(T^*\leq x)$ is the empirical cumulative distribution function of $T^*$. Then the  approximate $100(1-\eta)\%$ Boot-t confidence interval of $R$ is obtained by }
\begin{equation}
(\hat R_{Bt}(\frac{\eta}{2}),\hat R_{Bt}(1-\frac{\eta}{2})).\nonumber
\end{equation}
\end{itemize}

\section{Bayes Estimation of $R$}
In this section, {by assuming that the parameters $\mu$, $\lambda$ and $\alpha$ are random 
variables, the Bayesian inference of the unknown parameter $R$ is developed.} We study  the Bayes estimates and the associated credible intervals of $R$.  We use  the fact that  if $\mu$ is known, then $\lambda$ and $\alpha$ have gamma
conjugate priors  in our study.   If  all  three parameters are unknown,  then the joint conjugate priors do not exist.  In such a case , even for complete sample data,  all the elements of the expected Fisher information matrix are not finite. Therefore,  the Jeffrey's prior does not exist for this case.   We consider the following priors for $\lambda$, $\alpha$ and $\mu$ which  are fairly general.
  When $\mu$  is known, 
 $\lambda$ and $\alpha$ have the conjugate gamma priors. So we consider the following  priors for $\lambda$ and $\alpha$, 
$$\pi_1(\lambda)\propto\lambda^{a_1-1} e^{-b_1\lambda},\hspace{1cm}\lambda>0,a_1>0,b_1>0,$$
$$\pi_2(\alpha)\propto\alpha^{a_2-1} e^{-b_2\alpha},\hspace{1cm}\alpha>0,a_2>0,b_2>0.$$
Also the following non-proper  uniform prior is considered for $\mu$
$$\pi_3(\mu)\propto 1,\hspace{1cm}0<\mu<t_1.$$
Moreover, these random variables  are assumed to be independent.  So the joint posterior density function of $\lambda$, $\alpha$ and $\mu$, based on the observed sample, can be written  as
\begin{align}\label{L}
\pi(\lambda,\alpha,\mu|data)=\frac{L(data|\lambda,\alpha,\mu)\pi_1(\lambda)\pi_2(\alpha)\pi_3(\mu)}{\int_0^{t_1} \int_0^\infty \int_0^\infty L(data|\lambda,\alpha,\mu)\pi_1(\lambda)\pi_2(\alpha)\pi_3(\mu)d\lambda d\alpha d\mu}.
\end{align}
{The Bayes estimatos  cannot be obtained in a closed form by (\ref{L}).  So we adopt the Gibbs and Metropolis sampling techniques
 to compute the Bayes estimator and credible interval  of $R$.} The posterior pdfs of $\lambda$, $\alpha$ and $\mu$ 
can be written as:
$$\lambda|\alpha,\mu,\text{data}\sim\Gamma(n+a_1, b_1+\sum_{i=1}^{n}(R_i+1)(x_i-\mu)^2),$$
$$\alpha|\lambda,\mu,\text{data}\sim\Gamma(m+a_2, b_2+\sum_{j=1}^{m}(S_j+1)(y_j-\mu)^2),$$
and
\begin{align}
\pi(\mu|\lambda,\alpha,\text{data})&\propto e^{-\lambda b_1-\alpha b_2}\prod_{i=1}^{n}(x_i-\mu) \prod_{j=1}^{m}(y_j-\mu)
\nonumber\\
&\times \exp\left\{-\lambda\sum_{i=1}^{n}(R_i+1)(x_i-\mu)^2-\alpha\sum_{j=1}^{m}(S_j+1)(y_j-\mu)^2\right\}.\nonumber
\end{align}
\begin{theo}\label{theo1}
The conditional distribution of $\mu$ given $\lambda$, $\alpha$ and data is log-concave.
\end{theo}
\proof  The second derivative of the log conditional posterior
\begin{align}
\frac{\partial^2\ln(\pi(\mu|\lambda,\alpha,data))}{\partial\mu^2}&=-\left\{\sum_{i=1}^n\frac{1}{(x_i-\mu)^2}+\sum_{j=1}^m\frac{1}{(y_j-\mu)^2}\right.\nonumber\\
&+\left.2\lambda\sum_{i=1}^n(R_i+1)+2\alpha\sum_{j=1}^m(S_j+1)\right\}<0.\nonumber
\end{align}
So the conditional  posterior is log-concave. $\square$
\par
Devroye \cite{8fme} and Geman and Geman \cite{15fme} presented general methods to generate samples from a general log-concave density function and samples from the conditional posterior density functions, respectively. 
   So Theorem \ref{theo1} enables us to follow the method of Devroye \cite{8fme} and utilize  the idea of Geman and Geman \cite{15fme} to  generate $\mu_{(t)}$,   $\lambda_{(t)}$ and  $\alpha_{(t)}$ that using these samples we generate samples of $R$ that by which  we  provide estimates of  posterior mean and posterior  variance and   a $100(1-\eta)\%$ HPD credible interval for $R$.  
So we consider the following scheme: 
\begin{itemize}
\item 1. {Start the initial values ($\lambda_{(0)}$, $\alpha_{(0)}$, $\mu_{(0)}$) for ($\lambda$, $\alpha$, $\mu$).}
\item 2. Set $t=1$.
\item 3. Generate $\mu_{(t)}$ from $\pi(\mu|\lambda_{(t-1)},\alpha_{(t-1)}, \text{data})$.
\item 4. Generate $\lambda_{(t)}$ from $\Gamma(n+a_1, b_1+\sum\limits_{i=1}^{n}(R_i+1)(x_i-\mu_{(t-1)})^2)$.
\item 5. Generate $\alpha_{(t)}$ from $\Gamma(m+a_2, b_2+\sum\limits_{j=1}^{m}(S_j+1)(y_j-\mu_{(t-1)})^2)$.
\item 6. Compute $R_{(t)}$ from (\ref{R}).
\item 7. Set $t = t +1$.
\item 8. Repeat steps 3-7, $T$ times.
\end{itemize}
Then the estimates of  posterior mean and posterior variance of $R$ are evaluated as
$$\hspace{-.5in}\hat E(R|\text{data})=\frac{1}{T}\sum_{t=1}^{T}R_{(t)},\hspace{+.10in}
\hat Var(R|\text{data})=\frac{1}{T}\sum_{t=1}^{T}(R_{(t)}-\hat E(R|\text{data}))^2.$$
Applying the method of Chen and Shao \cite{10as},  a $100(1-\eta)\%$ HPD credible interval for $R$  is  obtained by $\left(R_{[\frac{\eta}{2}T]},R_{[(1-\frac{\eta}{2})T]}\right)$, where
$R_{[\frac{\eta}{2}T]}$ and $R_{[(1-\frac{\eta}{2})T]}$ are the $[\frac{\eta}{2}T]$-th and the $[(1-\frac{\eta}{2})T]$-th order statistics 
of $\{R_{(t)} : t=1,2,\ldots,T\}$, respectively.
\par
Now we consider Bayes estimation of $R$ when  $\lambda$ and $\alpha$ are random variables and  $\mu$ is known.  Assume that 
 $\lambda$ and $\alpha$ are independent  and have gamma priors with parameters $(a_1,b_1)$ and $(a_2,b_2)$, respectively. Then their  posterior density functions  are  $\Gamma(n+a_1, b_1+A_1(\bold x))$ and  $\Gamma(m+a_2, b_2+A_2(\bold y))$, respectively, where  $A_1(\bold x))=T_1/\lambda$ and $A_2(\bold y))=T_2/\alpha$. 
Thus  the posterior pdf of $R$  can be written as 
\begin{equation*}\label{post R} 
f_{R}(r)=S\frac{r^{m+a_2-1}(1-r)^{n+a_1-1}}{\left[r(b_2+A_2(\bold y))+(1-r)(b_1+A_1(\bold x))\right]^{n+m+a_1+a_2}},\hspace{1cm}0<r<1.
\end{equation*}
where
$$S=\frac{\Gamma(n+m+a_1+a_2)}{\Gamma(n+a_1)\Gamma(m+a_2)}(b_1+A_1(\bold x))^{n+a_1}(b_2+A_2(\bold y))^{m+a_2}.$$
\par
The Bayes estimate of $R$ under the squared error loss function can not be obtained in a closed form.  So following 
 the method of Lindley \cite{22sh} and  the approach of Ahmad et al. \cite{1sh}, we obtain the   approximate Bayes estimator  of $R$ under the squared error loss function as
\begin{equation}\label{R BS}
\hat R_{BS}=\tilde R\Bigg\{1+\frac{\tilde\lambda\tilde R^2}{\tilde\alpha^2(m+a_2-1)(n+b_1-1)}\left[\tilde\lambda(n+a_1-1)-\tilde\alpha(m+a_2-1)\right]\Bigg\},
\end{equation}
where
$$\tilde R=\frac{\tilde\alpha}{\tilde\alpha+\tilde\lambda},\hspace{1cm}\tilde\lambda=\frac{n+a_1-1}{b_1+A_1(\bold x)}, \hspace{1cm} \mbox{and} \hspace{1cm}\tilde\alpha=\frac{m+a_2-1}{b_2+A_2(\bold y)}.$$
The $100(1-\eta)\%$ Bayesian interval for $R$ is given by $(L,U)$, where 
\begin{equation}\label{CILind}
P[R\leq L|\mbox{data}]=\frac{\eta}{2}, \hspace{1cm}\mbox{and}\hspace{1cm} P[R\leq U|\mbox{data}]=1-\frac{\eta}{2}.
\end{equation}

\section{Data Analysis and Comparison Study}
In this section,  we compare the performance of the different estimators  and confidence intervals.  These comparison are   based on Monte Carlo simulations and real data experiments.
\subsection{Numerical experiments and discussions}
Here we compare  the performance of ML, UMVU and Bayes estimations by using Monte Carlo simulation.  These comparisons are  based on the  biases  and mean squares errors (MSE). Also,  the asymptotic, bootstrap and HPD  confidence intervals are compared based on average confidence lengths and coverage percentages. The bootstrap confidence intervals are evaluated based on 250 re-sampling. The Bayes estimates and the corresponding credible intervals are evaluated based on 1000 sampling,  $T = 1000$. Simulation are performed for different set of  parameters.   Also different sampling schemes are considered.   Comparing  of  the MLEs and  Bayes estimators are performed for three  sets of the parameters as  $\Theta_1=(\lambda=1, \alpha=1, \mu=1)$, $\Theta_2=(\lambda=1, \alpha=1,\mu=1.5)$ and $\Theta_3=(\lambda=1, \alpha=1,\mu=2.5)$.
Three priors are considered  for evaluating  Bayesian estimations and HPD credible intervals as: 
\begin{center}
Prior 1:\hspace{1cm} $a_j=0,$ \hspace{1cm} $b_j=0,$ \hspace{1cm} $j=1,2,$\\
Prior 2:\hspace{1cm} $a_j=1,$ \hspace{1cm} $b_j=1,$ \hspace{1cm} $j=1,2,$\\
Prior 3:\hspace{1cm} $a_j=2,$ \hspace{1cm} $b_j=3,$ \hspace{1cm} $j=1,2.$\\
\end{center}
Prior 1 is the non-informative gamma prior. Priors 2 and 3 are informative gamma priors.   We also consider  three censoring schemes (C.S.) which are explained in Table \ref{table1}.
\begin{table}[hb]
{\footnotesize
\caption{{\footnotesize Censoring schemes.\label{table1}}}
\begin{center}
\begin{tabular}{|c c c| }
\hline
& $(m,n)$ & C.S.\\
 \hline
 $r_1$ & (10,30) & (0,0,0,0,0,0,0,0,0,20)\\
 $r_2$ & (10,30) & (20,0,0,0,0,0,0,0,0,0)\\
 $r_3$ & (10,30) & (2,2,2,2,2,2,2,2,2,2)\\
\hline
\end{tabular}
\end{center}
}
\end{table}
\par
The average biases and MSEs of the MLEs and Bayes estimates  with different priors,  for different set of parameters and   censoring schemes,  with1000 replications  are reported in Table \ref{table2}.  Table \ref{table2} shows  that the MLE compares very well with the
Bayes estimators  in terms of biases and MSEs.  Also  Bayes estimator  with  informative gamma prior 3 clearly  outperform the one  with gamma  prior 2 in terms of both biases and MSEs. Moreover,  the Bayes estimators based on both gamma  priors outperform the one  obtained by the non-informative prior 1.

\begin{table}
{\footnotesize
\caption{{\footnotesize Biases and MSE of the MLE and Bayes estimates of $R$. \label{table2} }}
\begin{center}
\begin{tabular}{|c|ccccccccccccc|}
\hline
\multicolumn{1}{|c|}{$\Theta_j$} & \multicolumn{1}{c|}{C.S} & \multicolumn{3}{c|}{MLE} & \multicolumn{3}{c|}{Prior1} & \multicolumn{3}{c|}{Prior2} & \multicolumn{3}{c|}{Prior3} \\
\cline{3-14}
\multicolumn{1}{|c|}{} & \multicolumn{1}{c|}{} & \multicolumn{2}{c}{$\mid\mbox{Bias}\mid$} & \multicolumn{1}{c|}{MSE} & \multicolumn{2}{c}{$\mid\mbox{Bias}\mid$} & \multicolumn{1}{c|}{MSE} & \multicolumn{2}{c}{$\mid\mbox{Bias}\mid$} & \multicolumn{1}{c|}{MSE} & \multicolumn{2}{c}{$\mid\mbox{Bias}\mid$} & \multicolumn{1}{c|}{MSE} \\
\hline
\multicolumn{1}{|c|}{} & \multicolumn{1}{c|}{$(r_1,r_1)$} & \multicolumn{2}{c}{0.0061} & \multicolumn{1}{c|}{0.0204} & \multicolumn{2}{c}{0.0033} & \multicolumn{1}{c|}{0.0177} & \multicolumn{2}{c}{0.0019} & \multicolumn{1}{c|}{0.0143} & \multicolumn{2}{c}{0.0017} & \multicolumn{1}{c|}{0.0135} \\
\multicolumn{1}{|c|}{} & \multicolumn{1}{c|}{$(r_2,r_2)$} & \multicolumn{2}{c}{0.0021} & \multicolumn{1}{c|}{0.0187} & \multicolumn{2}{c}{0.0021} & \multicolumn{1}{c|}{0.0162} & \multicolumn{2}{c}{0.0010} & \multicolumn{1}{c|}{0.0140} & \multicolumn{2}{c}{0.0007} & \multicolumn{1}{c|}{0.0136} \\
\multicolumn{1}{|c|}{$\Theta_1$} & \multicolumn{1}{c|}{$(r_3,r_3)$} & \multicolumn{2}{c}{0.0095} & \multicolumn{1}{c|}{0.0194} & \multicolumn{2}{c}{0.0053} & \multicolumn{1}{c|}{0.0173} & \multicolumn{2}{c}{0.0035} & \multicolumn{1}{c|}{0.0136} & \multicolumn{2}{c}{0.0013} & \multicolumn{1}{c|}{0.0135} \\
\multicolumn{1}{|c|}{} & \multicolumn{1}{c|}{$(r_1,r_2)$} & \multicolumn{2}{c}{0.0246} & \multicolumn{1}{c|}{0.0165} & \multicolumn{2}{c}{0.0095} & \multicolumn{1}{c|}{0.0164} & \multicolumn{2}{c}{0.0022} & \multicolumn{1}{c|}{0.0158} & \multicolumn{2}{c}{0.0019} & \multicolumn{1}{c|}{0.0151} \\
\multicolumn{1}{|c|}{} & \multicolumn{1}{c|}{$(r_1,r_3)$} & \multicolumn{2}{c}{0.0188} & \multicolumn{1}{c|}{0.0176} & \multicolumn{2}{c}{0.0078} & \multicolumn{1}{c|}{0.0154} & \multicolumn{2}{c}{0.0069} & \multicolumn{1}{c|}{0.0129} & \multicolumn{2}{c}{0.0024} & \multicolumn{1}{c|}{0.0126} \\
\multicolumn{1}{|c|}{} & \multicolumn{1}{c|}{$(r_2,r_3)$} & \multicolumn{2}{c}{0.0120} & \multicolumn{1}{c|}{0.0180} & \multicolumn{2}{c}{0.0066} & \multicolumn{1}{c|}{0.0172} & \multicolumn{2}{c}{0.0025} & \multicolumn{1}{c|}{0.0170} & \multicolumn{2}{c}{0.0007} & \multicolumn{1}{c|}{0.0162} \\
\hline
\multicolumn{1}{|c|}{} & \multicolumn{1}{c|}{$(r_1,r_1)$} & \multicolumn{2}{c}{0.0122} & \multicolumn{1}{c|}{0.0176} & \multicolumn{2}{c}{0.0025} & \multicolumn{1}{c|}{0.0145} & \multicolumn{2}{c}{0.0022} & \multicolumn{1}{c|}{0.0136} & \multicolumn{2}{c}{0.0010} & \multicolumn{1}{c|}{0.0103} \\
\multicolumn{1}{|c|}{} & \multicolumn{1}{c|}{$(r_2,r_2)$} & \multicolumn{2}{c}{0.0083} & \multicolumn{1}{c|}{0.0164} & \multicolumn{2}{c}{0.0050} & \multicolumn{1}{c|}{0.0136} & \multicolumn{2}{c}{0.0039} & \multicolumn{1}{c|}{0.0135} & \multicolumn{2}{c}{0.0020} & \multicolumn{1}{c|}{0.0117} \\
\multicolumn{1}{|c|}{$\Theta_2$} & \multicolumn{1}{c|}{$(r_3,r_3)$} & \multicolumn{2}{c}{0.0038} & \multicolumn{1}{c|}{0.0167} & \multicolumn{2}{c}{0.0018} & \multicolumn{1}{c|}{0.0154} & \multicolumn{2}{c}{0.0017} & \multicolumn{1}{c|}{0.0154} & \multicolumn{2}{c}{0.0011} & \multicolumn{1}{c|}{0.0122} \\
\multicolumn{1}{|c|}{} & \multicolumn{1}{c|}{$(r_1,r_2)$} & \multicolumn{2}{c}{0.0109} & \multicolumn{1}{c|}{0.0163} & \multicolumn{2}{c}{0.0050} & \multicolumn{1}{c|}{0.0142} & \multicolumn{2}{c}{0.0020} & \multicolumn{1}{c|}{0.0133} & \multicolumn{2}{c}{0.0017} & \multicolumn{1}{c|}{0.0124} \\
\multicolumn{1}{|c|}{} & \multicolumn{1}{c|}{$(r_1,r_3)$} & \multicolumn{2}{c}{0.0131} & \multicolumn{1}{c|}{0.0176} & \multicolumn{2}{c}{0.0093} & \multicolumn{1}{c|}{0.0148} & \multicolumn{2}{c}{0.0087} & \multicolumn{1}{c|}{0.0129} & \multicolumn{2}{c}{0.0051} & \multicolumn{1}{c|}{0.0123} \\
\multicolumn{1}{|c|}{} & \multicolumn{1}{c|}{$(r_2,r_3)$} & \multicolumn{2}{c}{0.0141} & \multicolumn{1}{c|}{0.0192} & \multicolumn{2}{c}{0.0099} & \multicolumn{1}{c|}{0.0178} & \multicolumn{2}{c}{0.0049} & \multicolumn{1}{c|}{0.0165} & \multicolumn{2}{c}{0.0030} & \multicolumn{1}{c|}{0.0151} \\
\hline
\multicolumn{1}{|c|}{} & \multicolumn{1}{c|}{$(r_1,r_1)$} & \multicolumn{2}{c}{0.0150} & \multicolumn{1}{c|}{0.0152} & \multicolumn{2}{c}{0.0045} & \multicolumn{1}{c|}{0.0137} & \multicolumn{2}{c}{0.0012} & \multicolumn{1}{c|}{0.0124} & \multicolumn{2}{c}{0.0002} & \multicolumn{1}{c|}{0.0110} \\
\multicolumn{1}{|c|}{} & \multicolumn{1}{c|}{$(r_2,r_2)$} & \multicolumn{2}{c}{0.0110} & \multicolumn{1}{c|}{0.0143} & \multicolumn{2}{c}{0.0073} & \multicolumn{1}{c|}{0.0129} & \multicolumn{2}{c}{0.0027} & \multicolumn{1}{c|}{0.0111} & \multicolumn{2}{c}{0.0001} & \multicolumn{1}{c|}{0.0106} \\
\multicolumn{1}{|c|}{$\Theta_3$} & \multicolumn{1}{c|}{$(r_3,r_3)$} & \multicolumn{2}{c}{0.0096} & \multicolumn{1}{c|}{0.0130} & \multicolumn{2}{c}{0.0066} & \multicolumn{1}{c|}{0.0116} & \multicolumn{2}{c}{0.0052} & \multicolumn{1}{c|}{0.0114} & \multicolumn{2}{c}{0.0042} & \multicolumn{1}{c|}{0.0112} \\
\multicolumn{1}{|c|}{} & \multicolumn{1}{c|}{$(r_1,r_2)$} & \multicolumn{2}{c}{0.0120} & \multicolumn{1}{c|}{0.0135} & \multicolumn{2}{c}{0.0061} & \multicolumn{1}{c|}{0.0128} & \multicolumn{2}{c}{0.0037} & \multicolumn{1}{c|}{0.0126} & \multicolumn{2}{c}{0.0035} & \multicolumn{1}{c|}{0.0107} \\
\multicolumn{1}{|c|}{} & \multicolumn{1}{c|}{$(r_1,r_3)$} & \multicolumn{2}{c}{0.0174} & \multicolumn{1}{c|}{0.0168} & \multicolumn{2}{c}{0.0126} & \multicolumn{1}{c|}{0.0127} & \multicolumn{2}{c}{0.0086} & \multicolumn{1}{c|}{0.0114} & \multicolumn{2}{c}{0.0069} & \multicolumn{1}{c|}{0.0111} \\
\multicolumn{1}{|c|}{} & \multicolumn{1}{c|}{$(r_2,r_3)$} & \multicolumn{2}{c}{0.0139} & \multicolumn{1}{c|}{0.0137} & \multicolumn{2}{c}{0.0095} & \multicolumn{1}{c|}{0.0129} & \multicolumn{2}{c}{0.0085} & \multicolumn{1}{c|}{0.0122} & \multicolumn{2}{c}{0.0021} & \multicolumn{1}{c|}{0.0096} \\
\hline
\end{tabular}
\end{center}
}
\end{table}

\par
 Average length of the  $95\%$ asymptotic,  Boot-p and  Boot-t confidence intervals  and of  HPD credible intervals for different set of parameters and censoring schemes are presented 
 in Table \ref{table3}.    Results  in  Table \ref{table3} shows  that the bootstrap confidence intervals are wider than the other intervals.  The HPD credible intervals have the smallest average length for different censoring schemes and  different parameter values, and  the asymptotic confidence intervals are the second smallest  after HPD.  It is also observed that Boot-p confidence intervals have smaller average length  than the Boot-t confidence intervals. Also, it is evident that the Boot-t credible intervals provide the most coverage probabilities in most cases considered. Comparing the two HPD credible intervals based on the informative gamma priors clearly shows that the HPD credible intervals based on prior 3 have smaller  average length  than the HPD credible interval based on prior 2. The HPD credible intervals based on both priors have smaller average length than the ones obtained  by using the non-informative prior 1.

\begin{table}[h]
{\footnotesize
\caption{{\footnotesize Average confidence/credible length and coverage percentage for estimators of $R$. \label{table3} }}
\begin{center}
\begin{tabular}{ll|l|l|l|l|lll}
\hline
\multicolumn{9}{|c|}{$\Theta_1$} \\
\hline
\multicolumn{2}{|c|}{C.S} & \multicolumn{1}{c|}{MLE} & \multicolumn{1}{c|}{Boot-p} & \multicolumn{1}{c|}{Boot-t} & \multicolumn{4}{c|}{Bayes} \\
\cline{6-9}
\multicolumn{2}{|c|}{} & \multicolumn{1}{c|}{} & \multicolumn{1}{c|}{} & \multicolumn{1}{c|}{} & \multicolumn{1}{c|}{Prior1} & \multicolumn{1}{c|}{Prior2} & \multicolumn{2}{c|}{Prior3} \\
\hline
\multicolumn{2}{|c|}{$(r_1,r_1)$} & \multicolumn{1}{c|}{0.5144(0.935)} & \multicolumn{1}{c|}{0.5574(0.936)} & \multicolumn{1}{c|}{0.5757(0.949)} & \multicolumn{1}{c|}{0.4932(0.945)} & \multicolumn{1}{c|}{0.4558(0.942)} & \multicolumn{2}{c|}{0.3716(0.942)} \\
\multicolumn{2}{|c|}{$(r_2,r_2)$} & \multicolumn{1}{c|}{0.5023(0.932)} & \multicolumn{1}{c|}{0.5528(0.934)} & \multicolumn{1}{c|}{0.5883(0.950)} & \multicolumn{1}{c|}{0.4843(0.938)} & \multicolumn{1}{c|}{0.4406(0.947)} & \multicolumn{2}{c|}{0.3799(0.940)} \\
\multicolumn{2}{|c|}{$(r_3,r_3)$} & \multicolumn{1}{c|}{0.5169(0.934)} & \multicolumn{1}{c|}{0.5486(0.937)} & \multicolumn{1}{c|}{0.5822(0.952)} & \multicolumn{1}{c|}{0.4836(0.948)} & \multicolumn{1}{c|}{0.4485(0.946)} & \multicolumn{2}{c|}{0.3880(0.950)} \\
\multicolumn{2}{|c|}{$(r_1,r_2)$} & \multicolumn{1}{c|}{0.5143(0.947)} & \multicolumn{1}{c|}{0.5534(0.941)} & \multicolumn{1}{c|}{0.5837(0.955)} & \multicolumn{1}{c|}{0.4963(0.949)} & \multicolumn{1}{c|}{0.4586(0.940)} & \multicolumn{2}{c|}{0.4008(0.945)} \\
\multicolumn{2}{|c|}{$(r_1,r_3)$} & \multicolumn{1}{c|}{0.5193(0.942)} & \multicolumn{1}{c|}{0.5513(0.944)} & \multicolumn{1}{c|}{0.5911(0.958)} & \multicolumn{1}{c|}{0.4918(0.947)} & \multicolumn{1}{c|}{0.4466(0.942)} & \multicolumn{2}{c|}{0.3896(0.951)} \\
\multicolumn{2}{|c|}{$(r_2,r_3)$} & \multicolumn{1}{c|}{0.5146(0.930)} & \multicolumn{1}{c|}{0.5452(0.932)} & \multicolumn{1}{c|}{0.5944(0.953)} & \multicolumn{1}{c|}{0.4857(0.943)} & \multicolumn{1}{c|}{0.4469(0.946)} & \multicolumn{2}{c|}{0.3631(0.945)} \\
\hline
\multicolumn{9}{|c|}{$\Theta_2$} \\
\hline
\multicolumn{2}{|c|}{C.S} & \multicolumn{1}{c|}{MLE} & \multicolumn{1}{c|}{Boot-p} & \multicolumn{1}{c|}{Boot-t} & \multicolumn{4}{c|}{Bayes} \\
\cline{6-9}
\multicolumn{2}{|c|}{} & \multicolumn{1}{c|}{} & \multicolumn{1}{c|}{} & \multicolumn{1}{c|}{} & \multicolumn{1}{c|}{Prior1} & \multicolumn{1}{c|}{Prior2} & \multicolumn{2}{c|}{Prior3} \\
\hline
\multicolumn{2}{|c|}{$(r_1,r_1)$} & \multicolumn{1}{c|}{0.5145(0.946)} & \multicolumn{1}{c|}{0.5408(0.943)} & \multicolumn{1}{c|}{0.5947(0.959)} & \multicolumn{1}{c|}{0.4823(0.949)} & \multicolumn{1}{c|}{0.4547(0.944)} & \multicolumn{2}{c|}{0.3735(0.946)} \\
\multicolumn{2}{|c|}{$(r_2,r_2)$} & \multicolumn{1}{c|}{0.5142(0.942)} & \multicolumn{1}{c|}{0.5547(0.946)} & \multicolumn{1}{c|}{0.5841(0.950)} & \multicolumn{1}{c|}{0.4820(0.945)} & \multicolumn{1}{c|}{0.4510(0.944)} & \multicolumn{2}{c|}{0.3611(0.938)} \\
\multicolumn{2}{|c|}{$(r_3,r_3)$} & \multicolumn{1}{c|}{0.5200(0.948)} & \multicolumn{1}{c|}{0.5435(0.949)} & \multicolumn{1}{c|}{0.5931(0.959)} & \multicolumn{1}{c|}{0.4866(0.948)} & \multicolumn{1}{c|}{0.4494(0.946)} & \multicolumn{2}{c|}{0.3616(0.943)} \\
\multicolumn{2}{|c|}{$(r_1,r_2)$} & \multicolumn{1}{c|}{0.5100(0.947)} & \multicolumn{1}{c|}{0.5497(0.944)} & \multicolumn{1}{c|}{0.5883(0.958)} & \multicolumn{1}{c|}{0.4803(0.949)} & \multicolumn{1}{c|}{0.4526(0.943)} & \multicolumn{2}{c|}{0.3513(0.949)} \\
\multicolumn{2}{|c|}{$(r_1,r_3)$} & \multicolumn{1}{c|}{0.5156(0.946)} & \multicolumn{1}{c|}{0.5438(0.940)} & \multicolumn{1}{c|}{0.5875(0.957)} & \multicolumn{1}{c|}{0.4941(0.944)} & \multicolumn{1}{c|}{0.4574(0.948)} & \multicolumn{2}{c|}{0.3778(0.948)} \\
\multicolumn{2}{|c|}{$(r_2,r_3)$} & \multicolumn{1}{c|}{0.5133(0.945)} & \multicolumn{1}{c|}{0.5500(0.947)} & \multicolumn{1}{c|}{0.5852(0.956)} & \multicolumn{1}{c|}{0.4986(0.946)} & \multicolumn{1}{c|}{0.4557(0.952)} & \multicolumn{2}{c|}{0.3786(0.943)} \\
\hline
\multicolumn{9}{|c|}{$\Theta_3$} \\
\hline
\multicolumn{2}{|c|}{C.S} & \multicolumn{1}{c|}{MLE} & \multicolumn{1}{c|}{Boot-p} & \multicolumn{1}{c|}{Boot-t} & \multicolumn{4}{c|}{Bayes} \\
\cline{6-9}
\multicolumn{2}{|c|}{} & \multicolumn{1}{c|}{} & \multicolumn{1}{c|}{} & \multicolumn{1}{c|}{} & \multicolumn{1}{c|}{Prior1} & \multicolumn{1}{c|}{Prior2} & \multicolumn{2}{c|}{Prior3} \\
\hline
\multicolumn{2}{|c|}{$(r_1,r_1)$} & \multicolumn{1}{c|}{0.5023(0.935)} & \multicolumn{1}{c|}{0.5539(0.938)} & \multicolumn{1}{c|}{0.5920(.958)} & \multicolumn{1}{c|}{0.4962(0.947)} & \multicolumn{1}{c|}{0.4511(0.948)} & \multicolumn{2}{c|}{0.3760(0.951)} \\
\multicolumn{2}{|c|}{$(r_2,r_2)$} & \multicolumn{1}{c|}{0.5006(0.942)} & \multicolumn{1}{c|}{0.5511(0.943)} & \multicolumn{1}{c|}{0.5836(0.955)} & \multicolumn{1}{c|}{0.4975(0.950)} & \multicolumn{1}{c|}{0.4509(0.943)} & \multicolumn{2}{c|}{0.3612(0.943)} \\
\multicolumn{2}{|c|}{$(r_3,r_3)$} & \multicolumn{1}{c|}{0.5195(0.946)} & \multicolumn{1}{c|}{0.5527(0.945)} & \multicolumn{1}{c|}{0.5954(0.954)} & \multicolumn{1}{c|}{0.4823(0.941)} & \multicolumn{1}{c|}{0.4461(0.950)} & \multicolumn{2}{c|}{0.3705(0.944)} \\
\multicolumn{2}{|c|}{$(r_1,r_2)$} & \multicolumn{1}{c|}{0.5167(0.947)} & \multicolumn{1}{c|}{0.5538(0.951)} & \multicolumn{1}{c|}{0.5970(0.953)} & \multicolumn{1}{c|}{0.4858(0.950)} & \multicolumn{1}{c|}{0.4423(0.942)} & \multicolumn{2}{c|}{0.3693(0.943)} \\
\multicolumn{2}{|c|}{$(r_1,r_3)$} & \multicolumn{1}{c|}{0.5170(0.940)} & \multicolumn{1}{c|}{0.5591(0.945)} & \multicolumn{1}{c|}{0.5991(0.958)} & \multicolumn{1}{c|}{0.4883(0.947)} & \multicolumn{1}{c|}{0.4570(0.948)} & \multicolumn{2}{c|}{0.3784(0.943)} \\
\multicolumn{2}{|c|}{$(r_2,r_3)$} & \multicolumn{1}{c|}{0.5140(0.941)} & \multicolumn{1}{c|}{0.5527(0.941)} & \multicolumn{1}{c|}{0.5801(0.958)} & \multicolumn{1}{c|}{0.4889(0.948)} & \multicolumn{1}{c|}{0.4546(0.946)} & \multicolumn{2}{c|}{0.3664(0.947)} \\
\hline
\end{tabular}
\end{center}
}
\end{table}

\par Now we consider the case that the common location parameter $\mu$ is known.  So, utilizing (\ref{UMVUE}) and (\ref{R_known}) we obtain the {UMVU and ML estimates} of $R$ , respectively.  If there is  no prior information on $R$, the non-informative prior by assuming   $a_1 = b_1 = a_2 = b_2 = 0$  in  (\ref{R BS}) can be used to evaluate {the Lindley approximation of the Bayes estimates.}   Then the Bayesian interval  based on the Lindley approximation  can be derived  by (\ref{CILind}). 
 The average biases and MSEs 
of  the MLE, UMVUE and Bayes estimator based on 1000 replications are reported in Table \ref{table4}, where    
{ the  MLEs  are the best estimators as provide the smallest biases and MSEs, and the UMVUEs are the second best estimators.} 
Results of Tables \ref{table3} and \ref{table4} show that the  Bayesian intervals  based on Lindley approximation provide the smallest average credible lengths.  \\

\begin{table}[h]
{\footnotesize
\caption{{\footnotesize Biases and MSE of the MLE, UMVUE and Bayes estimators of $R$ and average confidence length and coverage percentage when $\mu$ is known. \label{table4} }}
\begin{center}
\begin{tabular}{ll|ll|ll|lllll}
\hline
\multicolumn{11}{|c|}{$\mu=0$} \\
\hline
\multicolumn{2}{|c|}{C.S} & \multicolumn{2}{c|}{MLE} & \multicolumn{2}{c|}{Lindley} & \multicolumn{2}{c|}{UMVUE} & \multicolumn{3}{c|}{Lindley} \\
\cline{3-8}
\multicolumn{2}{|c|}{} & \multicolumn{1}{c}{$\mid\mbox{Bias}\mid$} & \multicolumn{1}{c|}{MSE} & \multicolumn{1}{c}{$\mid\mbox{Bias}\mid$} & \multicolumn{1}{c|}{MSE} & \multicolumn{1}{c}{$\mid\mbox{Bias}\mid$} & \multicolumn{1}{c|}{MSE} & \multicolumn{3}{c|}{} \\
\hline
\multicolumn{2}{|c|}{$(r_1,r_1)$} & \multicolumn{1}{c}{0.0014} & \multicolumn{1}{c|}{0.0111} & \multicolumn{1}{c}{0.0039} & \multicolumn{1}{c|}{0.0124} & \multicolumn{1}{c}{0.0033} & \multicolumn{1}{c|}{0.0122} & \multicolumn{3}{c|}{0.3633(0.944)} \\
\multicolumn{2}{|c|}{$(r_2,r_2)$} & \multicolumn{1}{c}{0.0041} & \multicolumn{1}{c|}{0.0098} & \multicolumn{1}{c}{0.0107} & \multicolumn{1}{c|}{0.0161} & \multicolumn{1}{c}{0.0095} & \multicolumn{1}{c|}{0.0122} & \multicolumn{3}{c|}{0.3519(0.940)} \\
\multicolumn{2}{|c|}{$(r_3,r_3)$} & \multicolumn{1}{c}{0.0012} & \multicolumn{1}{c|}{0.0110} & \multicolumn{1}{c}{0.0048} & \multicolumn{1}{c|}{0.0142} & \multicolumn{1}{c}{0.0017} & \multicolumn{1}{c|}{0.0111} & \multicolumn{3}{c|}{0.3361(0.943)} \\
\multicolumn{2}{|c|}{$(r_1,r_2)$} & \multicolumn{1}{c}{0.0015} & \multicolumn{1}{c|}{0.0109} & \multicolumn{1}{c}{0.0088} & \multicolumn{1}{c|}{0.0122} & \multicolumn{1}{c}{0.0031} & \multicolumn{1}{c|}{0.0115} & \multicolumn{3}{c|}{0.3411(0.948)} \\
\multicolumn{2}{|c|}{$(r_1,r_3)$} & \multicolumn{1}{c}{0.0019} & \multicolumn{1}{c|}{0.0111} & \multicolumn{1}{c}{0.0082} & \multicolumn{1}{c|}{0.0121} & \multicolumn{1}{c}{0.0040} & \multicolumn{1}{c|}{0.0110} & \multicolumn{3}{c|}{0.3470(0.949)} \\
\multicolumn{2}{|c|}{$(r_2,r_3)$} & \multicolumn{1}{c}{0.0025} & \multicolumn{1}{c|}{0.0104} & \multicolumn{1}{c}{0.0050} & \multicolumn{1}{c|}{0.0177} & \multicolumn{1}{c}{0.0029} & \multicolumn{1}{c|}{0.0120} & \multicolumn{3}{c|}{0.3515(0.950)} \\
\hline
\multicolumn{11}{|c|}{$\mu=1$} \\
\hline
\multicolumn{2}{|c|}{C.S} & \multicolumn{2}{c|}{MLE} & \multicolumn{2}{c|}{Lindley} & \multicolumn{2}{c|}{UMVUE} & \multicolumn{3}{c|}{Lindley} \\
\cline{3-8}
\multicolumn{2}{|c|}{} & \multicolumn{1}{c}{$\mid\mbox{Bias}\mid$} & \multicolumn{1}{c|}{MSE} & \multicolumn{1}{c}{$\mid\mbox{Bias}\mid$} & \multicolumn{1}{c|}{MSE} & \multicolumn{1}{c}{$\mid\mbox{Bias}\mid$} & \multicolumn{1}{c|}{MSE} & \multicolumn{3}{c|}{} \\
\hline
\multicolumn{2}{|c|}{$(r_1,r_1)$} & \multicolumn{1}{c}{0.0024} & \multicolumn{1}{c|}{0.0105} & \multicolumn{1}{c}{0.0030} & \multicolumn{1}{c|}{0.0130} & \multicolumn{1}{c}{0.0025} & \multicolumn{1}{c|}{0.0127} & \multicolumn{3}{c|}{0.3475(0.945)} \\
\multicolumn{2}{|c|}{$(r_2,r_2)$} & \multicolumn{1}{c}{0.0017} & \multicolumn{1}{c|}{0.0106} & \multicolumn{1}{c}{0.0052} & \multicolumn{1}{c|}{0.0116} & \multicolumn{1}{c}{0.0021} & \multicolumn{1}{c|}{0.0110} & \multicolumn{3}{c|}{0.3447(0.946)} \\
\multicolumn{2}{|c|}{$(r_3,r_3)$} & \multicolumn{1}{c}{0.0045} & \multicolumn{1}{c|}{0.0113} & \multicolumn{1}{c}{0.0054} & \multicolumn{1}{c|}{0.0137} & \multicolumn{1}{c}{0.0050} & \multicolumn{1}{c|}{0.0123} & \multicolumn{3}{c|}{0.3395(0.940)} \\
\multicolumn{2}{|c|}{$(r_1,r_2)$} & \multicolumn{1}{c}{0.0013} & \multicolumn{1}{c|}{0.0104} & \multicolumn{1}{c}{0.0021} & \multicolumn{1}{c|}{0.0126} & \multicolumn{1}{c}{0.0012} & \multicolumn{1}{c|}{0.0123} & \multicolumn{3}{c|}{0.3380(0.955)} \\
\multicolumn{2}{|c|}{$(r_1,r_3)$} & \multicolumn{1}{c}{0.0027} & \multicolumn{1}{c|}{0.0106} & \multicolumn{1}{c}{0.0057} & \multicolumn{1}{c|}{0.0154} & \multicolumn{1}{c}{0.0046} & \multicolumn{1}{c|}{0.0119} & \multicolumn{3}{c|}{0.3354(0.949)} \\
\multicolumn{2}{|c|}{$(r_2,r_3)$} & \multicolumn{1}{c}{0.0030} & \multicolumn{1}{c|}{0.0105} & \multicolumn{1}{c}{0.0071} & \multicolumn{1}{c|}{0.0121} & \multicolumn{1}{c}{0.0044} & \multicolumn{1}{c|}{0.0115} & \multicolumn{3}{c|}{0.3405(0.950)} \\
\hline
\multicolumn{11}{|c|}{$\mu=1.5$} \\
\hline
\multicolumn{2}{|c|}{C.S} & \multicolumn{2}{c|}{MLE} & \multicolumn{2}{c|}{Lindley} & \multicolumn{2}{c|}{UMVUE} & \multicolumn{3}{c|}{Lindley} \\
\cline{3-8}
\multicolumn{2}{|c|}{} & \multicolumn{1}{c}{$\mid\mbox{Bias}\mid$} & \multicolumn{1}{c|}{MSE} & \multicolumn{1}{c}{$\mid\mbox{Bias}\mid$} & \multicolumn{1}{c|}{MSE} & \multicolumn{1}{c}{$\mid\mbox{Bias}\mid$} & \multicolumn{1}{c|}{MSE} & \multicolumn{3}{c|}{} \\
\hline
\multicolumn{2}{|c|}{$(r_1,r_1)$} & \multicolumn{1}{c}{0.0011} & \multicolumn{1}{c|}{0.0106} & \multicolumn{1}{c}{0.0066} & \multicolumn{1}{c|}{0.0124} & \multicolumn{1}{c}{0.0018} & \multicolumn{1}{c|}{0.0109} & \multicolumn{3}{c|}{0.3380(0.947)} \\
\multicolumn{2}{|c|}{$(r_2,r_2)$} & \multicolumn{1}{c}{0.0023} & \multicolumn{1}{c|}{0.0105} & \multicolumn{1}{c}{0.0037} & \multicolumn{1}{c|}{0.0121} & \multicolumn{1}{c}{0.0033} & \multicolumn{1}{c|}{0.0108} & \multicolumn{3}{c|}{0.3498(0.952)} \\
\multicolumn{2}{|c|}{$(r_3,r_3)$} & \multicolumn{1}{c}{0.0015} & \multicolumn{1}{c|}{0.0109} & \multicolumn{1}{c}{0.0022} & \multicolumn{1}{c|}{0.0127} & \multicolumn{1}{c}{0.0018} & \multicolumn{1}{c|}{0.0121} & \multicolumn{3}{c|}{0.3395(0.947)} \\
\multicolumn{2}{|c|}{$(r_1,r_2)$} & \multicolumn{1}{c}{0.0016} & \multicolumn{1}{c|}{0.0104} & \multicolumn{1}{c}{0.0025} & \multicolumn{1}{c|}{0.0113} & \multicolumn{1}{c}{0.0013} & \multicolumn{1}{c|}{0.0110} & \multicolumn{3}{c|}{0.3401(0.943)} \\
\multicolumn{2}{|c|}{$(r_1,r_3)$} & \multicolumn{1}{c}{0.0016} & \multicolumn{1}{c|}{0.0101} & \multicolumn{1}{c}{0.0030} & \multicolumn{1}{c|}{0.0155} & \multicolumn{1}{c}{0.0023} & \multicolumn{1}{c|}{0.0116} & \multicolumn{3}{c|}{0.3432(0.947)} \\
\multicolumn{2}{|c|}{$(r_2,r_3)$} & \multicolumn{1}{c}{0.0015} & \multicolumn{1}{c|}{0.0107} & \multicolumn{1}{c}{0.0038} & \multicolumn{1}{c|}{0.0142} & \multicolumn{1}{c}{0.0017} & \multicolumn{1}{c|}{0.0121} & \multicolumn{3}{c|}{0.3446(0.946)}  \\
\hline
\end{tabular}
\end{center}
}
\end{table}

\subsection{Data analysis}
{ Here  we consider  the strength data  which reported by Badar and Priest \cite{2sh}.}   These data  represent the strength measured in GPA for single carbon fibers and impregnated $1000$-carbon fiber tows. Single fibers were tested under tension at gauge lengths of 20mm (Data Set $1$) and 10mm (Data Set $2$).
The data are presented in Tables \ref{table5} and \ref{table6}.  We have subtracted $0.75$  from all the data of  both these data sets. 
\par
\begin{figure}
\begin{center}
\vspace{.5cm}
 \includegraphics[width=13cm, height=6cm,  angle=0] {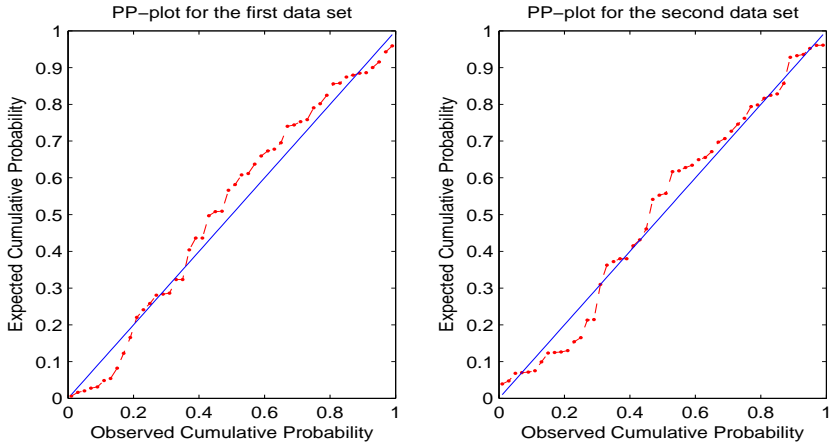}
\end{center}
\vspace{-.75cm}
\caption{\footnotesize{The PP-plots for $X$ and $Y$}.\label{Fig1}}
\end{figure}

{We  check the  fitness of  two-parameter Rayleigh distribution for  the data sets, separately.  For the first data set  parameters are estimated as $\hat\lambda=2.6708,\;\hat\mu=1.9188$  and for the second data set as $\hat\alpha=1.0349,\;\hat\mu=2.2574$.} The Kolmogorov-Smirnov distances are $0.0881$ and $0.0967$ and corresponding p-values are $0.8004$ and $0.7013$, respectively.  The  p-p plots are presented  in Figure \ref{Fig1}. {The p-values indicate that the two-parameter Rayleigh distributions provide  adequate fit for these data sets.}
\par
Based on the complete data set the MLE estimation of $R$ is 0.1270 and the {associated} $95\%$ confidence interval is  (0.0668,0.2476). The Bayes estimate of $R$ with respect to {non-informative} priors is 0.1373 and the {corresponding} 95$\%$ credible interval is (0.0619,0.2603).\par
We have considered  two different progressively censored samples  from the above data sets, where the corresponding censored schemes are presented in Table \ref{table7}. Based on Scheme 1, the MLE and Bayes estimates are 0.1820 and 0.1852, respectively. The associated $95\%$ asymptotic confidence interval and credible interval are (0.0101,0.2033) and (0.0208,0.2261), respectively.   Based on Scheme 2, the MLE and Bayes estimates are 0.1682 and 0.1566, and the associated $95\%$ asymptotic confidence interval and  credible interval are (0.0030,0.2083) and (0.0272,0.2145), respectively.  Clearly, the estimates obtained based on Scheme 2  are closer to the estimates obtained by complete sample.

\begin{table}
{\footnotesize
\caption{{\footnotesize Data Set 1 (gauge lengths of 20 mm). \label{table5}}}
\begin{center}
\begin{tabular}{|c c c c c c c c c c| }
\hline
 1.966 & 1.997 & 2.006 & 2.021 & 2.027 & 2.055 & 2.063 & 2.098 & 2.140 & 2.179 \\
 2.224 & 2.240 & 2.253 & 2.270 & 2.272 & 2.274 & 2.301 & 2.301 & 2.359 & 2.382 \\
2.382 & 2.426 & 2.434 & 2.435 & 2.478 & 2.490 & 2.511 & 2.514 & 2.535 & 2.554 \\
 2.566 & 2.570 & 2.586 & 2.629 & 2.633 & 2.642 & 2.648 & 2.684 & 2.697 & 2.726 \\
2.770 & 2.773 & 2.800 & 2.809 & 2.818 & 2.821 & 2.848 & 2.880 & 2.954 & 3.012 \\
\hline
\end{tabular}
\end{center}
}
\end{table}
\vspace{-.5cm}
\begin{table}
{\footnotesize
\caption{{\footnotesize Data Set 2 (gauge lengths of 10 mm). \label{table6}}}
\begin{center}
\begin{tabular}{|c c c c c c c c c c|}
\hline
2.454 & 2.474 & 2.518 & 2.522 & 2.525 & 2.532 & 2.575 & 2.614 & 2.616 & 2.618 \\
 2.624 & 2.659 & 2.675 & 2.738 & 2.740 & 2.856 & 2.917 & 2.928 & 2.937 & 2.937 \\
2.977 & 2.996 & 3.030 & 3.125 & 3.139 & 3.145 & 3.220 & 3.223 & 3.235 & 3.243 \\
 3.264 & 3.272 & 3.294 & 3.332 & 3.346 & 3.377 & 3.408 & 3.435 & 3.493 & 3.501 \\
3.537 & 3.554 & 3.562 & 3.628 & 3.852 & 3.871 & 3.886 & 3.971 & 4.024 & 4.027  \\
 \hline
\end{tabular}
\end{center}
}
\end{table}

\begin{table}
{\footnotesize
\caption{{\footnotesize Censored schemes and the corresponding data. \label{table7}}}
\begin{center}
\begin{tabular}{|ccccccccccc|}
\hline
\multicolumn{11}{|c|}{Scheme1: $R=S=[4,4,4,4,4,4,4,4,4,4]$} \\
\hline
$i,j$ & 1 & 2 & 3 & 4 & 5 & 6 & 7 & 8 & 9 & 10 \\
$x_i$ & 1.966 & 2.055 & 2.224 & 2.274 & 2.382 & 2.490 & 2.566 & 2.642 & 2.770 & 2.821 \\
$y_j$ & 2.454 & 2.532 & 2.624 & 2.856 & 2.977 & 3.145 & 3.264 & 3.377 & 3.537 & 3.871 \\
\hline
\multicolumn{11}{|c|}{Scheme2: $R=S=[2,2,2,2,2,2,2,2,2,22]$} \\
\hline
 $i,j$& 1 & 2 & 3 & 4 & 5 & 6 & 7 & 8 & 9 & 10 \\
$x_i$ & 1.966 & 2.021 & 2.063 & 2.179 & 2.253 & 2.274 & 2.359 & 2.426 & 2.478 & 2.514 \\
$y_j$ & 2.454 & 2.522 & 2.575 & 2.618 & 2.675 & 2.856 & 2.937 & 2.996 & 3.139 & 3.223 \\
\hline
\end{tabular}
\end{center}
}
\end{table}

\section{Conclusions}
In this paper, the estimation of the stress-strength parameter for two-parameter Rayleigh distribution under progressive Type-II censoring scheme is {studied.   For the case that location parameters are known, the exact confidence interval of $R$ is obtained.   Assuming that the location parameters are equal but unknown, different methods for the estimating of $R=P(Y<X)$ are utilized.  As the MLE of $R$ can not be obtained analytically, an iterative procedure is applied to compute it.  Moreover,  the asymptotic confidence interval is derived by using the observed Fisher information matrix.}  It is observed that even for small  sample sizes  
 the asymptotic confidence intervals work quite well.  Also, two bootstrap confidence intervals were proposed that their performance is quite satisfactory. {Using the Gibbs sampling, the Bayes estimate of $R$ and  corresponding credible interval are 
 obtained  too. The simulation results show that} the MLE compares very well with the Bayes estimator in terms of biases and MSEs. {Assuming that the  location parameter is known}, MLE, UMVUE and  Bayes estimators are evaluated.  The  MLE provides the smallest biases and MSEs and the UMVUEs are {the second best} estimators.  Finally,  Bayesian intervals  based on Lindley approximation provide the smallest average credible lengths in compare to other methods.

\end{document}